\documentclass{ws-procs9x6}

\begin{document}

\title{ON GENERALIZED OSCILLATOR ALGEBRAS AND THEIR ASSOCIATED COHERENT STATES}

\author{J. D. BUKWELI-KYEMBA$^\dagger$ and M. N. HOUNKONNOU$^*$ }

\address{International Chair in Mathematical Physics and  Applications\\
(ICMPA - UNESCO Chair),
University of Abomey-Calavi\\
Cotonou, 072 B.P.: 50, Republic of Benin\\
$^\dagger$E-mail:desbuk@gmail.com\\
$^*$E-mail: norbert.hounkonnou@cipma.uac.bj; hounkonnou@yahoo.fr}

\begin{abstract}
A unified method of calculating structure functions  from commutation relations of deformed single-mode oscillator algebras 
is  presented. A natural approach to building coherent states associated to  deformed algebras is then deduced.
\end{abstract}


\bodymatter
\section{Introduction}

Generalization of  Heisenberg-Weyl algebra was suggested long before the discovery of quantum groups by Heisenberg to achieve regularization for (nonrenormalizable) nonlinear spinor field theory. The issue was considered as small additions to the canonical commutation relations~\cite{Dancoff,Tamm}. 
Snyder~\cite{Snyder}, investigating the infrared catastrophe of soft photons in the Compton scattering, raised this issue   and built a noncommutative Lorentz invariant space-time where the noncommutativity of space operators is proportional to nonlinear combinations of phase space operators. 
 
Further modifications of the oscillator algebra and their possible physical interpretations as spectrum generating algebras 
of non standard statistics were made since the earlier work of Snyder.  As matter of citation, 
let us mention the $q$-oscillator algebras by  Coon and coworkers~\cite{Arik&Coon,Yu&al72}, Kuryshkin~\cite{Kuryshkin80}, 
Jannussis and  collaborators~\cite{Jannussis&al83b}.         

With the development  of quantum groups, new aspects of $q$-oscillators were identified~\cite{Biedenharn,Macfarlane,Sun&Fu} as 
a tool for providing a boson realization of  quantum algebra $su_q(2)$ using a $q$-analogue of the harmonic oscillator and 
the Jordan-Schwinger mapping, and then generalizations in view of unifying  or extending different existing $q$-deformed algebras
 were elaborated~\cite{Burban1,Meljanac}

Despite all useful properties and applications motivated by various one-parameter deformed algebras, 
 the multi-parameter deformations aroused much interest because of their  flexibility 
 when  dealing with concrete physical models. See for example~\cite{Baloitcha,Hounkonnou&Bukweli10} and references therein. 

Coherent states (CS) have practically followed the same trend as the quantum algebras. They were invented
 by Schr\"odinger in 1926 in  the context of the quantum harmonic oscillator, and defined as minimum-uncertainty 
states that exhibit a classical behavior~\cite{Schrodinger}.

In 1963, coherent states were simultaneous rediscovered by Glauber~\cite{Glauber1,Glauber2}, Klauder~\cite{Klauder63a,Klauder63b} and Sudarshan~\cite{Sudarshan} in   quantum optics of coherent light beams emitted by lasers.

The vast field covered  by coherent states motivated their generalizations to other families of states deducible from noncanonical operators and satisfying not necessarily all  properties of CS~\cite{Ali&al99,Perelomov}.

The more general class is essentially based on the overcompleteness property of coherent states. This property was the~{\it raison d'\^etre} of the mathematically oriented construction of generalized coherent states by Ali~{\it et al} \cite{Ali&al99}
 or of the ones with physical orientations~\cite{Gazeau09,Klauder&al}.      
  Numerous publications continue to appear  using this property. See for example~\cite{Daoud,Hounkonnou&Ngompe07b}
 and references therein.

In this paper we give a method of computing the so-called structure function which is the basis
 of coherent state construction  from a given algebra. 

\section{Unified deformed single-mode oscillator algebras and associated CS}\label{Sec2}

The Heisenberg-Weyl algebra is a Lie algebra generated by the position $Q$, the momentum $P$ and the unity $\mathbf{1}$ operators satisfying:
\begin{equation}
 [Q,\;P]= i\hbar\mathbf{1},\quad [Q,\;\mathbf{1}]=0,\quad[P,\;\mathbf{1}]=0.
\end{equation}
Defining the annihilation and creation operators by
\begin{math}
 b:=\frac{Q+iP}{\sqrt{2\hbar}}\;\mbox{and}\; b^\dagger:=\frac{Q-iP}{\sqrt{2\hbar}},
\end{math}
the Weyl-Heisenberg algebra is generated by $\{b,\;b^\dagger,\;\mathbf{1}\}$ satisfying:
\begin{eqnarray}\label{Harmonic1}
 [b,\;b^\dagger]=\mathbf{1},\quad [b,\;\mathbf{1}]=0,\quad[b^\dagger,\;\mathbf{1}]=0.
\end{eqnarray}
From (\ref{Harmonic1}), one defines the operator  $N:=b^\dagger b$, also called {\it number operator}, with the properties:
\begin{eqnarray}\label{Harmonic2}
 [N,\;b]=-b,\quad [N,\;b^\dagger]=b^\dagger,\quad [N,\;\mathbf{1}]=0.
\end{eqnarray}

The canonical coherent states (CS) are normalized states $|z\rangle$ satisfying one of the  following three equivalent conditions \cite{Glauber1,Klauder63a,Schrodinger,Sudarshan}:
\begin{eqnarray*}
&&(i)\quad(\Delta Q)(\Delta P)= \frac{\hbar}{2},\cr
&&(ii)\quad b|z\rangle= z|z\rangle;\cr
&&(iii)\quad |z\rangle=e^{zb^\dagger-\bar z b}|0\rangle= e^{-|z|^2/2}e^{zb^\dagger}|0\rangle=e^{-|z|^2/2}\sum_{n=0}^\infty\frac{z^n}{\sqrt{n!}}|n\rangle,
\end{eqnarray*}
where $(\Delta A)^2:=\langle z|A^2-\langle A\rangle^2|z\rangle,\; \langle A\rangle:= \langle z|A|z\rangle$, $b|0\rangle= 0$ and $\langle 0|0\rangle=1$.

They belong to the Fock space $\mathcal{F}=\mathrm{span}\{|n\rangle\;|\;n\in \mathbb{N}\cup \{0\}\}$, where the states
\begin{eqnarray}
 |n\rangle=\frac{1}{\sqrt{n!}}(b^\dagger)^n|0\rangle,\;\; n= 1,\;2,\;\cdots
\end{eqnarray}
satisfy the orthogonality and completeness conditions:
\begin{eqnarray}
 \langle m|n\rangle=\delta_{m,n}, \;\quad \sum_{n=0}^\infty |n\rangle \langle n|= \mathbf{1}.
\end{eqnarray}
The important feature of these coherent states resides in the partition (resolution) of identity:
\begin{eqnarray}
\int_\mathbb{C}{{[d^2z]}\over\pi} |z\rangle\langle z|= \mathbf{1},
\end{eqnarray}
where we have put $[d^2z]= d(Rez)d(Imz)$ for simplicity.
\begin{definition}
We  call deformed Heisenberg  algebra,  an associative algebra generated by the set of operators $\{\mathbf{1},\; a,\; a^\dagger,\; N\}$
satisfying the relations
\begin{eqnarray}\label{dal}
 [N,\;a^\dagger]= a^\dagger,\qquad [N,\;a]= -a,\label{uq01}
\end{eqnarray}
such that there exists a non-negative analytic function $f$, called the structure function,  defining the operator products $ a^\dagger a$ and $aa^\dagger$ in the following way:
\begin{eqnarray}
 a^\dagger a := f(N),\qquad aa^\dagger:= f(N+\mathbf{1}),\label{uq02}
\end{eqnarray}
where $N$ is a self-adjoint operator, $a$ and its Hermitian conjugate $a^\dagger$ denote the deformed annihilation and creation operators, respectively.
\end{definition}

The deformed Heisenberg algebras have a common property characterized by the existence of
a self-adjoint  number operator $N$, a lowering operator $a$  and its formal adjoint, called raising operator, $a^\dagger$, and  differ by the expression of the  structure function $f$.

The associated Fock  space $\mathcal{F}$ is now spanned by the orthonormalized eigenstates of the number operator $N$ given by:
\begin{eqnarray}
 |n\rangle= \frac{1}{\sqrt{f(n)!}}(a^\dagger)^n|0\rangle,\quad n\in\mathbb{N}\cup\{0\},
\end{eqnarray}
where
\begin{math}
f(n)!=f(n)f(n-1)...f(1)\;\mbox{ with}\; f(0)=0.
\end{math}
Moreover,
\begin{eqnarray}
 a|n\rangle= \sqrt{f(n)}|n-1\rangle,\quad a^\dagger|n\rangle=\sqrt{f(n+1)}|n+1\rangle.
\end{eqnarray}
In this context, the structure function $f$ appears as a key unifying methods of coherent state construction corresponding to deformed algebras.

Denote $\mathbf{D}_f:=\left\{z\in\mathbb{C}:|z|^2<R_f\right\}$, where $R_f$ is the radius of convergence of the series (the {\it deformed exponential function}):
\begin{eqnarray}\label{ser1}
\mathcal{N}_f(x):=\sum_{n=0}^\infty \frac{x^n}{[f(n)]!}.
\end{eqnarray}
Then, the  states
\begin{eqnarray}
 |z,f\rangle 
&:=&(\mathcal{N}_f(|z|^2))^{-1/2}\sum_{n=0}^\infty \frac{z^n}{\sqrt{[f(n)]!}}|n\rangle,\;\;z\in\mathbf{D}_f,
\end{eqnarray}
are normalized eigenstates of  the raising operator $ a$ with the eigenvalues $z$. They are not orthogonal to each other.
Moreover, the map $z\mapsto|z,f\rangle$ from $\mathbf{D}_f\subset\mathbb{C}$ to the Fock space $\mathcal{F}$ is continuous.

The set $\left\{|z,f\rangle: z\in\mathbf{D}_f\right\}$ will be called {\it family of coherent states}  whether there exists a positive measure $\mu_f$ such that~\cite{Klauder&Skagerstam}
\begin{eqnarray}\label{unitf}
 \int_{\mathbf{D}_f} d\mu_f(\bar z,z)|z,f\rangle\langle z,f|= \sum_{n=0}^\infty|n\rangle\langle n|=\mathbf{1}.
\end{eqnarray}
Passing to polar coordinates,  $z=\sqrt{x} e^{i\theta}$, 
where $0\leq\theta\leq 2\pi$, $ 0<x< R_f$, and $d\mu(\bar z,z)= d\omega_f(x)d\theta$, this corresponds to the following classical
power-moment problem \cite{Akhiezer,Tarmakin}:
\begin{eqnarray}
\label{PMoment}
 \int_0^{R_f}x^n\;\frac{2\pi\;d\omega_f(x)}{\mathcal{N}_f(x)}=[f(n)]!,\quad n= 0,\; 1,\; 2,\; ...
\end{eqnarray}
Note immediately that not  all deformed algebras lead to coherent states in this construction formalism because the moment problem (\ref{PMoment}) does  not always have a solution~\cite{Akhiezer,Tarmakin}.

Many techniques have been proposed in the literature to determine the structure function $f$ corresponding to a given algebra~\cite{Baloitcha,Burban1,Meljanac}.

Meljanac~{\it et al}~\cite{Meljanac} defined the generalized $q$-deformed single-mode oscillator as an algebra generated by the identity operator $\mathbf{1}$, a self-adjoint number operator $N$,  a lowering operator $a$ and  an  operator $\bar{a}$ which is not necessarily conjugate to $a$ satisfying
\begin{eqnarray}
 && [N,\;a]= -a, \quad 
[N,\;\bar{a}]= \bar{a}, \label{uq2}\\
&& a\bar{a}-F(N)\bar{a}a=G(N)  \label{uq3}
\end{eqnarray}
where $F$ and $G$ are arbitrary complex analytic functions.

Such an algebra furnishes an appropriate approach for the  unification of  classes of  deformed algebras known in the literature.

For the purpose, let us  start from the relations (\ref{uq2}) to get
\begin{math}
[N,\;a\bar{a}]= 0=[N,\;\bar{a}a]
\end{math}
implying the existence of a complex analytic function $\varphi$ such that
\begin{eqnarray}
 \bar{a}a=\varphi(N)\quad\mbox{and}\qquad a\bar{a}=\varphi(N+1).\label{uq4}
\end{eqnarray}
Eqs (\ref{uq3}) and (\ref{uq4}) give
\begin{equation}
 \varphi(N+1)-F(N)\varphi(N)= G(N).\label{uq5}
\end{equation}
Denote now $a^\dagger$ the Hermitian conjugate of the operator $a$. Then,
\begin{eqnarray}\label{uq7}
[N,\;a^\dagger]= a^\dagger,\qquad\mbox{and}\qquad \bar{a}=c(N)a^\dagger,
\end{eqnarray}
where $c(N)$ is a complex function. For convenience take $c(N)= e^{i\arg{\varphi(N)}}$. Therefore, from (\ref{uq4}) and the fact that $a^\dagger a$ and $aa^\dagger$ are Hermitian operators we necessarily have
\begin{eqnarray}\label{uq8}
 a^\dagger a= |\varphi(N)|\qquad\mbox{and}\qquad  aa^\dagger= |\varphi(N+1)|.
\end{eqnarray}
We now assume the existence of a "vacuum state" $|0\rangle$ such that 
\begin{equation}
 N|0\rangle=0,\; a|0\rangle=0 \; \mbox{and}\;
\langle 0|0\rangle= 1,
\end{equation}
and construct the non normalized eigenvectors, $(a^\dagger)^n|0\rangle$ of the operator $N$. Then, applying (\ref{uq5}) to these vectors we obtain
\begin{eqnarray} \label{stctr}
\varphi(n)= [F(n-1)]!\sum_{k=0}^{n-1}\frac{G(k)}{[F(k)]!},\quad n\geq 1,
\end{eqnarray}
unless the initial condition $\varphi(0)=0$ is satisfied, and where
\begin{eqnarray}
 [F(k)]!=\left\{\begin{array}{lcr}F(k)F(k-1)\cdots F(1)&\mbox{ if }& k\geq 1\\1&\mbox{ if }& k=0 \end{array}\right..
\end{eqnarray}
Hence,  the structure function characterizing a given deformation is defined as follows:
$f(n):= \varphi(n)$ if $\varphi(n)\geq0$, and $f(n):= |\varphi(n)|$, in general. Provided the function $f,$ the above formalism can
 be displayed to construct generalized CS.

\section*{Acknowledgments}
 This work is partially supported by the Abdus Salam International
 Centre for Theoretical Physics (ICTP, Trieste, Italy) through the
 Office of External Activities (OEA) - \mbox{Prj-15}. The ICMPA is in partnership with  the Daniel Iagolnitzer Foundation (DIF), France.

\bibliographystyle{ws-procs9x6}
\bibliography{Hounk-contr}

\end{document}